\documentclass[namedreferences]{solarphysics}
\usepackage[hyperref,optionalrh,solaromanenum,showbiblabels]{spr-sola-addons} 
\usepackage{graphicx}                    
\usepackage{amssymb}                    
\usepackage{color}                       
\usepackage{breakurl}                         
\usepackage{threeparttable}

\chardef\us=`\_

\begin{document}

\begin{article}

\begin{opening}

\title{On the relationship between G-band bright point dynamics and their magnetic field strengths}

%
\author[addressref={aff1,aff2},email={yangyf@escience.cn}]{\inits{Y.F.}\fnm{Yunfei}~\lnm{Yang}}
\author[addressref=aff1]{\inits{Q.}\fnm{Qiang}~\lnm{Li}}
\author[addressref={aff1}, corref, email={jikaifan@cnlab.net}]{\inits{K.F.}\fnm{Kaifan}~\lnm{Ji}}
\author[addressref={aff1,aff2}]{\inits{S.}\fnm{Song}~\lnm{Feng}}
\author[addressref=aff1]{\inits{H.}\fnm{Hui}~\lnm{Deng}}
\author[addressref={aff1,aff3}]{\inits{F.}\fnm{Feng}~\lnm{Wang}}
\author[addressref={aff2}]{\inits{J.B.}\fnm{Jiaben}~\lnm{Lin}}

%
\runningauthor{Yang et al.}
\runningtitle{On the relationship between GBP dynamics and the magnetic field strengths}

\address[id={aff1}]{Faculty of Information Engineering and Automation / Yunnan Key Laboratory of Computer Technology Application,
Kunming University of Science and Technology, 650500, China}
\address[id={aff2}]{Key Laboratory of Solar Activity, National Astronomical Observatories, Chinese Academy of Sciences, 100012, China}
\address[id={aff3}]{Yunnan Observatory, Chinese Academy of Sciences, 650011, China}

\begin{abstract}
G-band bright points (GBPs) are regarded as good manifestations of magnetic flux concentrations. We aim to investigate the relationship between the dynamic properties of GBPs and their longitudinal magnetic field strengths. High spatial and temporal resolution observations were recorded simultaneously with G-band filtergrams and \textit{Narrow-band Filter Imager} (NFI) Stokes $I$ and $V$ images with \textit{Hinode} /\textit{Solar Optical Telescope}. The GBPs are identified and tracked in the G-band images automatically, and the corresponding longitudinal magnetic field strength of each GBP is extracted from the calibrated NFI magnetograms by a point-to-point method. After categorizing the GBPs into five groups by their longitudinal magnetic field strengths, we analyze the dynamics of GBPs of each group. The results suggest that with increasing longitudinal magnetic field strengths of GBPs correspond to a decrease in their horizontal velocities and motion ranges as well as by showing more complicated motion paths. This suggests that magnetic elements showing weaker magnetic field strengths prefer to move faster and farther along straighter paths, while stronger ones move more slowly in more erratic paths within a smaller region. The dynamic behaviors of GBPs with different longitudinal magnetic field strengths can be explained by that the stronger flux concentrations withstand the convective flows much better than weaker ones.
\end{abstract}

%

\end{opening}

%

\section{Introduction}
\label{sec:Introduction}
Bright points within photospheric intergranular dark lanes are easily visible signatures of small-scale magnetic flux tube concentrations \citep{Stenflo1985Measurements,Solanki1993Smallscale}, especially in G-band observations \citep{Steiner2001Radiative}, and so are often called as G-band bright points (GBPs). GBPs are jostled continuously by the evolution of surrounding granules, and herded into the dark downflow lanes between the granulation cells \citep{Berger2001On}. Their horizontal motions are thought to be the source of guiding magneto-hydrodynamic waves, which could contribute to coronal heating and solar wind expansion \citep{Roberts1983Wave,Choudhuri1993Implications,Cranmer2015role}. Thus, the dynamics of GBPs is a major interest when attempting to understand how the corona is heated and how the solar wind is accelerated.

GBPs have a mean horizontal velocity of 1--2\,$\rm km$ $\rm s^{-1}$ with the maximum value of 7\,$\rm km$ $\rm s^{-1}$ \citep{Muller1994proper,Berger1998Measurements,Nisenson2003Motions,Mostl2006Dynamics,Utz2010Dynamics,Keys2011Velocity,Yang2014Evolution}. In last decades, some authors compared the horizontal velocities of GBPs in different magnetized environments, such as quiet Sun (QS) regions and active regions (ARs). There is a general agreement that these results can be interpreted such as that the horizontal velocity of GBPs in ARs is attenuated compared to those in QS regions, in spite of different instruments and methods \citep{Berger1998Measurements,Mostl2006Dynamics,Keys2011Velocity,Criscuoli2014statistical}. Recently, \citet{Keys2014Dynamic} defined one QS sub-region and two active sub-regions in the same field of view (FOV) judging by their mean magnetic flux densities, and proposed that the horizontal velocity of GBPs in the QS sub-region is $\sim$30\% greater than that in the active sub-regions.

Besides that, some authors focused on tracking the rotary or vortex motion of GBPs, which is suggested as a primary candidate mechanism, for energy transport from the interior to the outer layers of the solar atmosphere \citep{Vogler2005Simulations,Bonet2008Convectively,Carlsson2010Chromospheric,Fedun2011magnetohydrodynamics,Moll2011Vortices,Shelyag2011Vorticity,Moll2012Vortices,Shelyag2013Alfven}. \citet{Yang2015Characterizing} proposed a path classifying method, and suggested that the motion paths of GBPs could be categorized into three types: straight, rotary, and erratic. In addition, the motion range of GBPs was analyzed and suggested that the majority of GBPs move in a local place \citep{Bodnarova2014On,Yang2015Characterizing}.

However, the study about the dynamics of GBPs with different magnetic field strengths is very lacking. This might be due to a lack of co-temporal high spatial and temporal resolution magnetic field strength data and G-band data wherefore only a few authors (e.g., \citealp{Beck2007Magnetic,Bovelet2008quiet,Viticchie2010Imaging,Utz2013Magnetic,Riethmuller2014Comparison}) tried to shed light on the magnetic field strength distribution of GBPs. The \textit{Hinode} /\textit{Solar Optical Telescope} (SOT; \citealp{Ichimoto2008Polarization,Suematsu2008Solar}) minimizes the deleterious effects of bad seeing conditions and image reconstruction from ground-based observations. It produces \textit{Narrow-band Filter Imager} (NFI) Stokes $I$ and $V$ images, which are co-spatial and co-temporal with the high spatial and temporal resolution \textit{Broad-band Filter Imager} (BFI) G-band images, and keep a large FOV in observations. These data provide a good opportunity to extract the simultaneous longitudinal magnetic field strength of each GBP during its lifetime.

In this article, we aim to investigate the relationship between the dynamics of GBPs and their longitudinal magnetic field strengths, in terms of the horizontal velocity, the motion range and the type of motion path. It will enlighten how fast, how far and how intricately the small-scale magnetic elements with different magnetic field strengths move on the solar surface. We will learn more about the interaction between magnetic fields and convection, and then further the knowledge of coronal heating and solar wind expansion. Not only that, the results might help modeling of MHD wave excitation and propagation simulations, \textit{e.g.}, \citet{Vigeesh2009Wave} suggested that the energy carried to the higher atmosphere by different driving of the flux tube in the photosphere changes the propagated energy flux by magnitudes, but they might be also needed as necessary input to perform more realistic simulations of the wave heating processes in the solar atmosphere.

The article is constructed as follows. The data and data analysis are described in Section~\ref{sec:DATA}. In Section~\ref{sec:RESULT}, the relationship between the dynamics of GBPs and their longitudinal magnetic field strengths is presented, followed by a discussion and the conclusions in Section~\ref{sec:DISCUSSION} and Section~\ref{sec:CONCLUSION}, respectively.

\section{Data and Data Analysis}
\label{sec:DATA}
\begin{figure}[ht]
\centerline{
\includegraphics[width=0.458\linewidth]{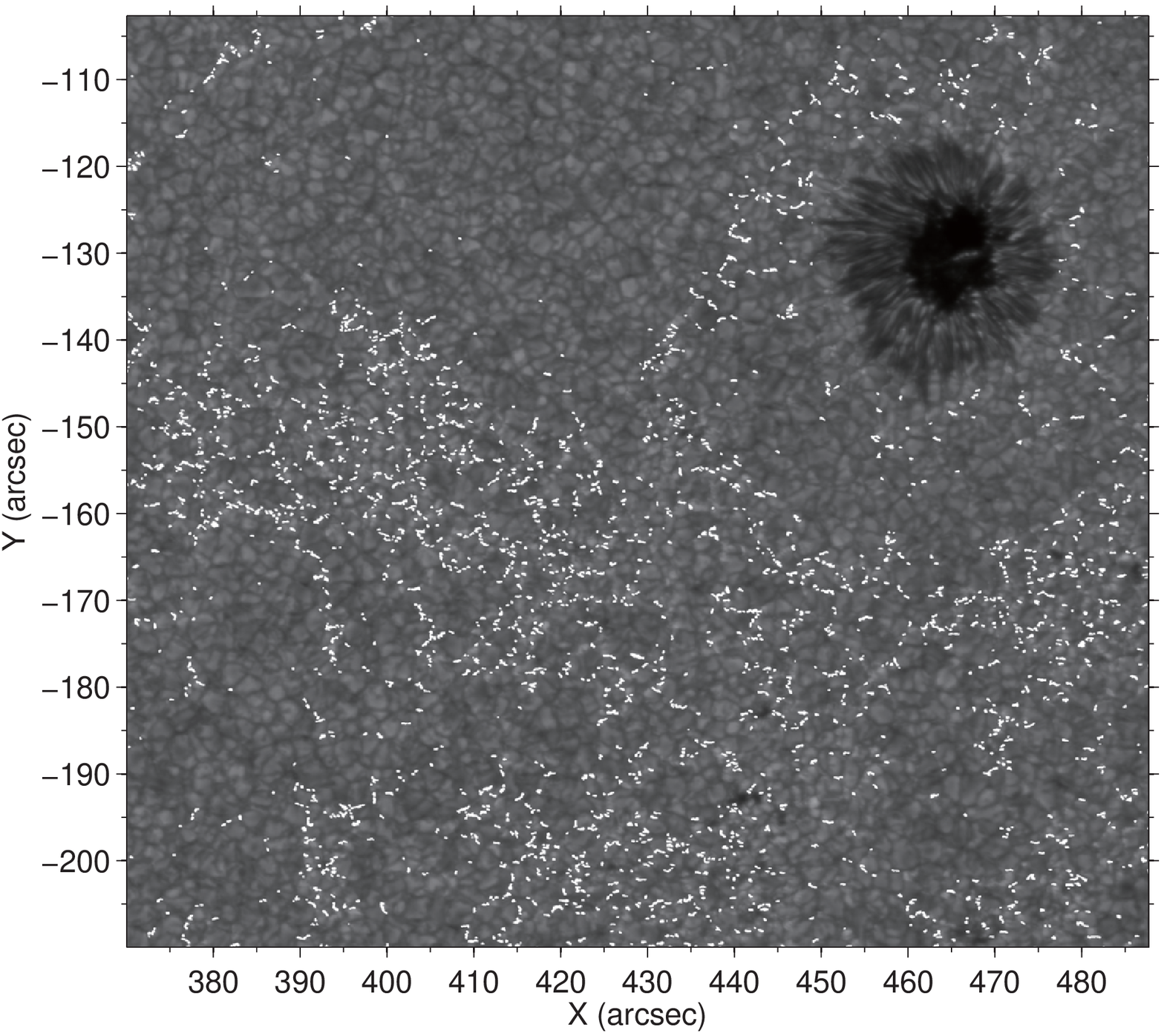}
\\ \\ \\
\includegraphics[width=0.52\linewidth]{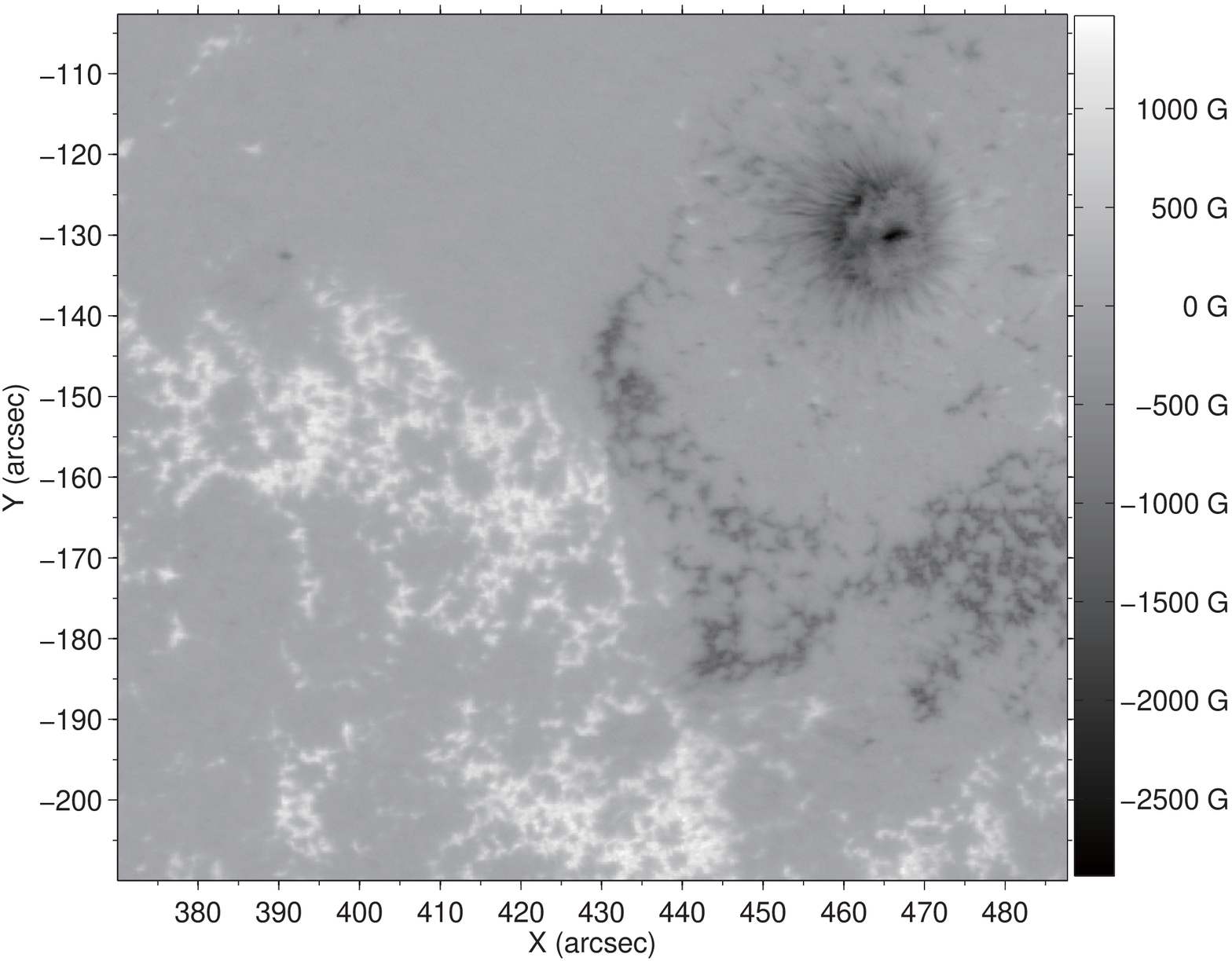}}
\caption{Left panel: the first frame of the G-band images, which was recorded at 22:31:21\,$\rm UT$ on 2007 June 9. The detected GBPs are highlighted with white contours. Right panel: the corresponding calibrated NFI magnetogram, which was recorded at 22:31:32\,$\rm UT$ on 2007 June 9. The $x$ and $y$ coordinates are in arcsec.}
    \label{fig:figure1}
\end{figure}

We studied a data set of NOAA 10960 acquired with \textit{Hinode} /SOT comprising BFI G-band filtergrams and NFI Stokes $I$ and $V$ images. G-band data with wavelengths of 430.5 $\rm nm$ give a better contrast between photospheric bright points and the surrounding granulations \citep{Steiner2001Radiative,Schussler2003Why}. The G-band time-series consists of 132 images taken between 22:31:21 and 23:37:26\,$\rm UT$ on 2007 June 9 with a cadence of 30\,$\rm s$. The spatial sampling is 0.109$^{\prime\prime}$ pixel$^{-1}$ over a 111.6$^{\prime\prime}$$\times$111.6$^{\prime\prime}$ FOV. The center of FOV is (429.5$^{\prime\prime}$, -158.5$^{\prime\prime}$) corresponding to a heliocentric angle of 28.2$^{\circ}$ equaling a cosine value of 0.90. The Stokes $I$ and $V$ images were taken 200 m{\AA} blueward of the Na \textsc{i} D 589.6 $\rm nm$ spectral line, which measures the photospheric longitudinal magnetic field. The Stokes $I$ and $V$ time-series is co-spatial and co-temporal with the G-band time-series, and the sampling time is delayed 11\,$\rm s$ compared to the G-band time-series. Its spatial sampling is 0.16$^{\prime\prime}$ pixel$^{-1}$ and cadence is 30\,$\rm s$. These FG data were calibrated and reduced using a standard data reduction algorithm fg\_prep.pro distributed in solar software. Figure~\ref{fig:figure1} shows the first frame of the G-band image-series, in which the detected GBPs are highlighted with white contours, and the corresponding NFI magnetogram.

First, with a reference to the SOT /spectro-polarimeter (SP) longitudinal magnetogram that was multiplied by the filling factors, we calibrated the NFI Stokes $V/I$ signals outside the sunspots in the same way as previous works \citep{Jefferies1989Transfer,Chae2007Initial,Ichimoto2008Polarization,Zhou2010Solar,Bai2013Calibration,Bai2014Improved,Yang2015Dispersal}. The mean longitudinal magnetic field strength (excluding the sunspots region) of the NFI calibrated magnetograms is 132\,$\rm G$, and the residual standard deviation is about 130\,$\rm G$. In addition, we selected a relatively quiet region (about 1$''\times$1$''$) in the NFI magnetograms to quantify the noise of NFI magnetograms, and got a standard deviation ($\sigma$) of 10\,$\rm G$.

Then, the temporally closest G-band images and NFI images were chosen. The different spatial samplings were overcome by bicubic interpolations of the NFI images obtaining adjusted NFI images with the spatial sampling of the G-band images. Next, the two time-series were aligned carefully by a high-accuracy solar image registration procedure based on a cross-correlation technique \citep{Yang2015Characterizing}: all G-band images in the time-series were aligned to its first image, and the NFI magnetograms were aligned to the G-band images according to the displacement between the simultaneous Stokes $I$ image and the G-band image. To estimate the error of the stretched NFI magnetograms, we have performed the following experiment. We stretched the NFI magnetogram, $I_{0}$, to a new magnetograms, $I_{1}$, which corresponds to the spatial sampling of the G-band data by bicubic interpolation. Then $I_{1}$ was down-sampled to $I_{2}$ that corresponds to the spatial sampling of the NFI magnetograms by bicubic interpolation, too. The mean absolute error between $I_{0}$ and $I_{2}$ is 2.88\,$\rm G$. The absolute error of the magnetic field strength peaks in the magnetogram maps (1103\,$\rm G$) is 25\,$\rm G$ and the relative error value is 2.3\%. Since the noise of NFI magnetograms is about 10\,$\rm G$, we obtain an error of the stretched NFI magnetograms of 10.40 G using an error transfer formula as $\sqrt{(\Delta B_{1})^{2}+(\Delta B_{2})^{2}}$, where $\Delta B_{1}$ and $\Delta B_{2}$ are the errors of the independent random variables $B_{1}$ and $B_{2}$. It shows that the error of the stretched NFI magnetogram is smaller than its noise. Besides that, we also measured the difference between every two sequential NFI magnetograms in this time-series (the sampling time is 30\,$\rm s$) to estimate the temporal difference between the G-band images and NFI magnetograms (the sampling time is delayed 11\,$\rm s$). The average difference is 0.03\,$\rm G$. This means that the difference of magnetograms in 11\,$\rm s$ would not be larger than 0.03\,$\rm G$. The value is much less than the noise of NFI magnetograms.

After that, we detected GBPs in each G-band image by a Laplacian and morphological dilation algorithm \citep{Feng2013Statistical}, and tracked the evolution of each GBP in the image-series by a three dimensional (3D) segment algorithm \citep{Yang2014Evolution}.
Some GBPs that may be identified wrongly due to detection errors were discarded. Among those GBPs are the ones with diameters of less than 100\,$\rm km$, or lifetimes shorter than 60\,$\rm s$, or horizontal velocities exceeding 7\,$\rm km s^{-1}$, or incomplete life cylces. Additionally, non-isolated GBPs, which merge or split during their lifetimes, were also discarded. As a result, a total of 103 023 isolated GBPs remain in 132 images, yielding 18 494 evolving GBPs.

The longitudinal magnetic field strength of each GBP is in principle easy to extract from the corresponding NFI magnetogram as the region of each GBP is known. However, considering the small-scale of GBPs even slightest errors in the image alignment process could degrade the average or median value of the detected MBP significantly \citep{Berger2007Contrast}. Thus we measured and used the peak unsigned longitudinal magnetic field strength of each GBP in this work. This value will be from now on abbreviated within this text as $B_P$. The presented and used method is also known as ``point-to-point".

At last, the geometric projection effects were corrected by multiplying the $x$ and $y$ coordinates after the identification of GBPs by the corresponding cosine and sine foreshortening factors.

\section{Results}
\label{sec:RESULT}

\subsection{The magnetic field strength distribution of GBPs}
\begin{figure}[ht]
\centerline{
\includegraphics[width=0.45\textwidth]{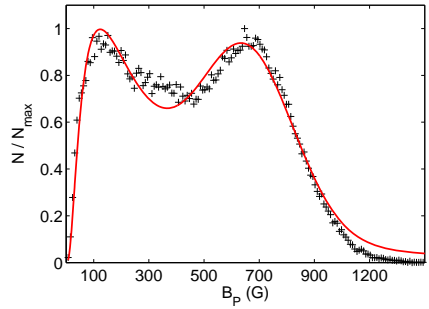}
}
\caption{The longitudinal magnetic field strength distribution of GBPs, which is fitted to a double log-normal distribution (red solid line). The y-axis is the normalized frequency ($N/N_{max}$, where $N$ is the number of each bin, $N_{max}$ is the maximum number of $N$).}
    \label{fig:figure2}
\end{figure}

Figure~\ref{fig:figure2} shows the $B_P$ distribution of the GBPs, which is fitted to a double log-normal distribution. The fit equation is:
\begin{equation}\label{eqa1}
\frac{w_1}{x\sigma_1\sqrt{2\pi}}e(^{-\frac{1}{2}(\frac{log(x)-\mu_1}{\sigma_1})^{2}})+\frac{w_2}{x\sigma_2\sqrt{2\pi}}e(^{-\frac{1}{2}(\frac{log(x)-\mu_2}{\sigma_2})^{2}}),
\end{equation}
where $w_1$ and $w_2$ specify the coefficients that give the relative contribution of each component to the total function, $\mu_1$ and $\mu_2$ are more readily treated for the geometric mean values ($e^\mu$), and $\sigma_1$ and $\sigma_2$ for the geometric standard deviations ($e^\sigma$) of the two components. The values of the parameters we derived are listed in Table~\ref{Tab:tab1}. The relative contributions of the two components are 67\% and 33\% respectively. Two peaks are at about 140 and 650\,$\rm G$, respectively. The log-normal distribution occurs naturally by physical mechanisms involved in generating of the magnetic field, which the process of fragmentation and merging dominate the process of magnetic flux concentration \citep{Abramenko2005Distribution}.

\begin{table}
\begin{center}
\caption[]{The parameters of the double log-normal distribution}\label{Tab:tab1}
  \begin{tabular}{cccccc}
  \hline\noalign{\smallskip}
  $w1$ & $w2$ &$\mu1$ &$\mu2$& $\sigma1$ & $\sigma2$  \\
  \hline\noalign{\smallskip}
      531$\pm$46  &259$\pm$28	&5.81$\pm$0.10 &6.52$\pm$0.01 &0.95$\pm$0.05 &0.21$\pm$0.02 \\
  \noalign{\smallskip}\hline
\end{tabular}
\end{center}
\end{table}

Taking the noise (3$\sigma$, 30\,$\rm G$) and the residual standard deviation (130\,$\rm G$) of the NFI magnetograms into account, the GBPs whose $B_P$ values range from 50 to 1050\,$\rm G$, which account for 97\% of all GBPs, were categorized into five groups by setting the bin as 200\,$\rm G$. Verbally, the group from 50 to 250\,$\rm G$ (not including 250) represents very weakly magnetized GBPs, from 250 to 450\,$\rm G$ (not including 450) represents weakly magnetized GBPs, from 450 to 650\,$\rm G$ (not including 650) represents magnetized GBPs, from 650 to 850\,$\rm G$ (not including 850) represents strongly magnetized GBPs, and from 850 to 1050\,$\rm G$ represents very strongly magnetized GBPs.

\subsection{The relation between $B_P$ and horizontal velocity}

\begin{figure}
\centering
\includegraphics[width=0.4\textwidth]{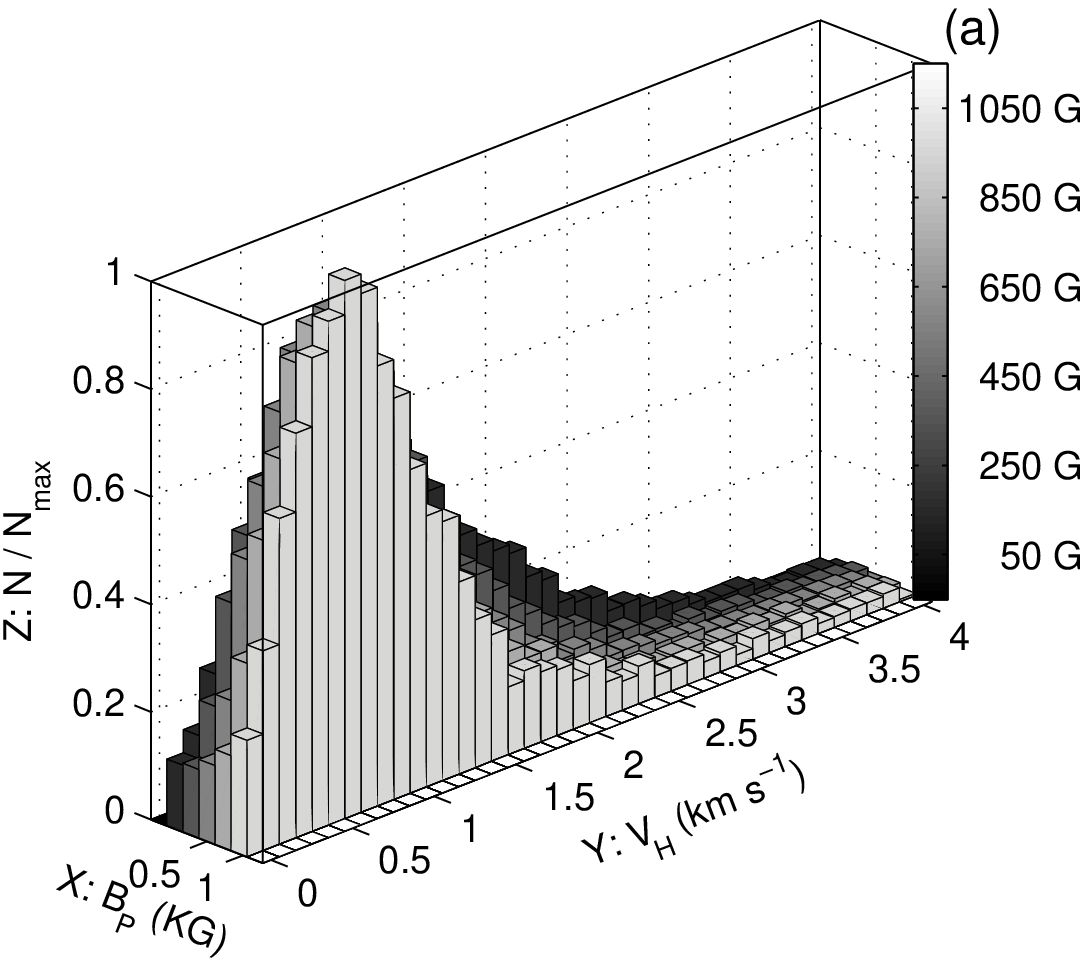}
\includegraphics[width=0.46\textwidth]{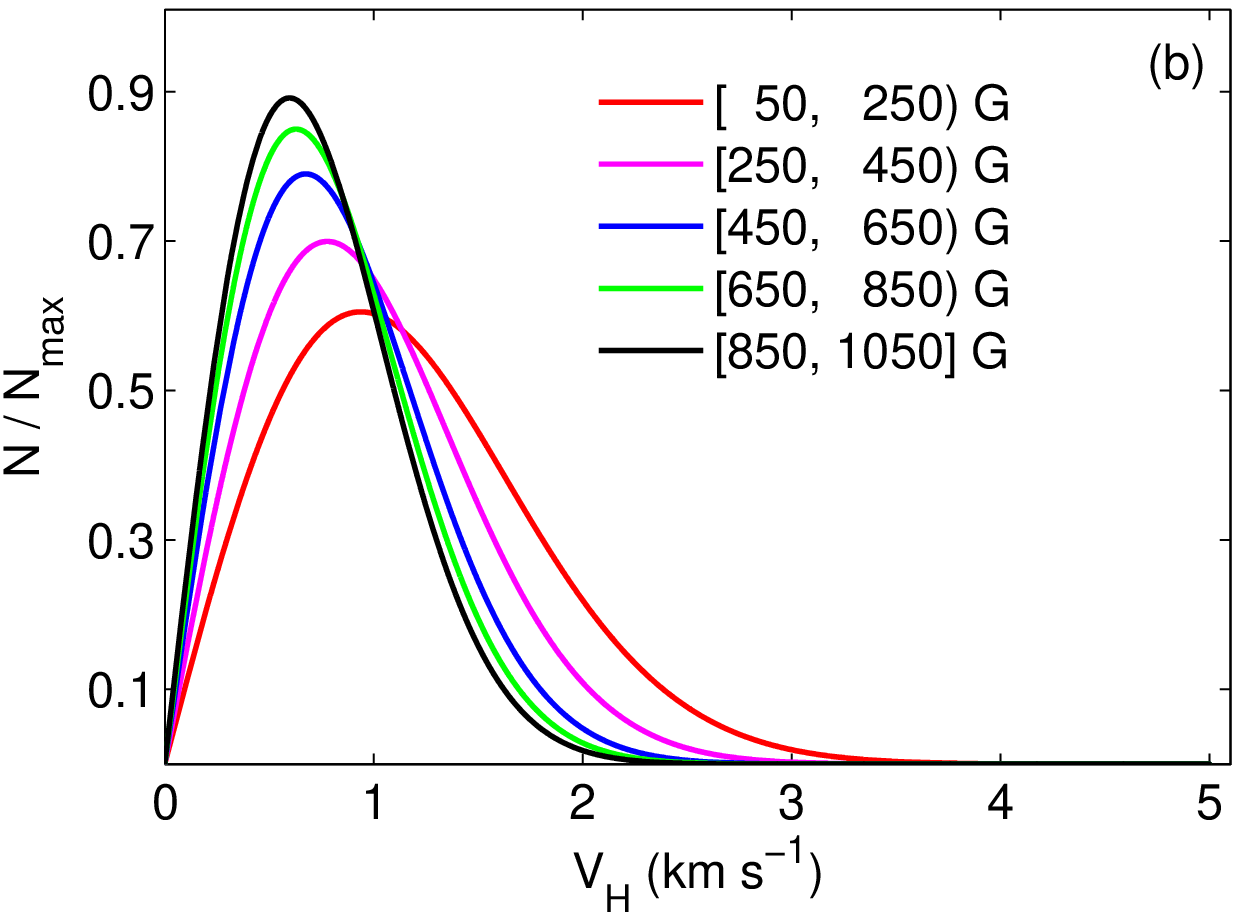}
\includegraphics[width=0.5\textwidth]{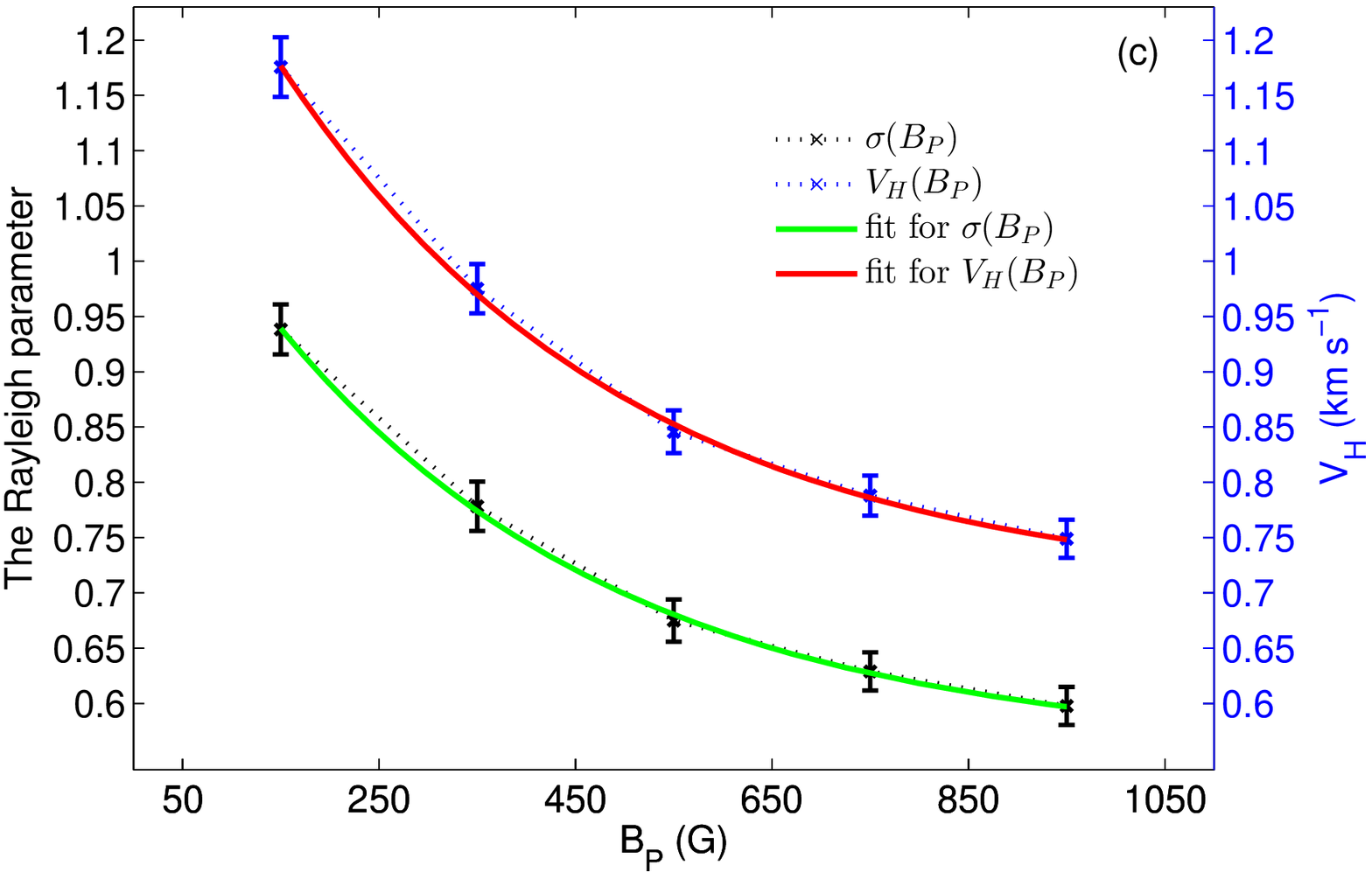}
\includegraphics[width=0.44\textwidth]{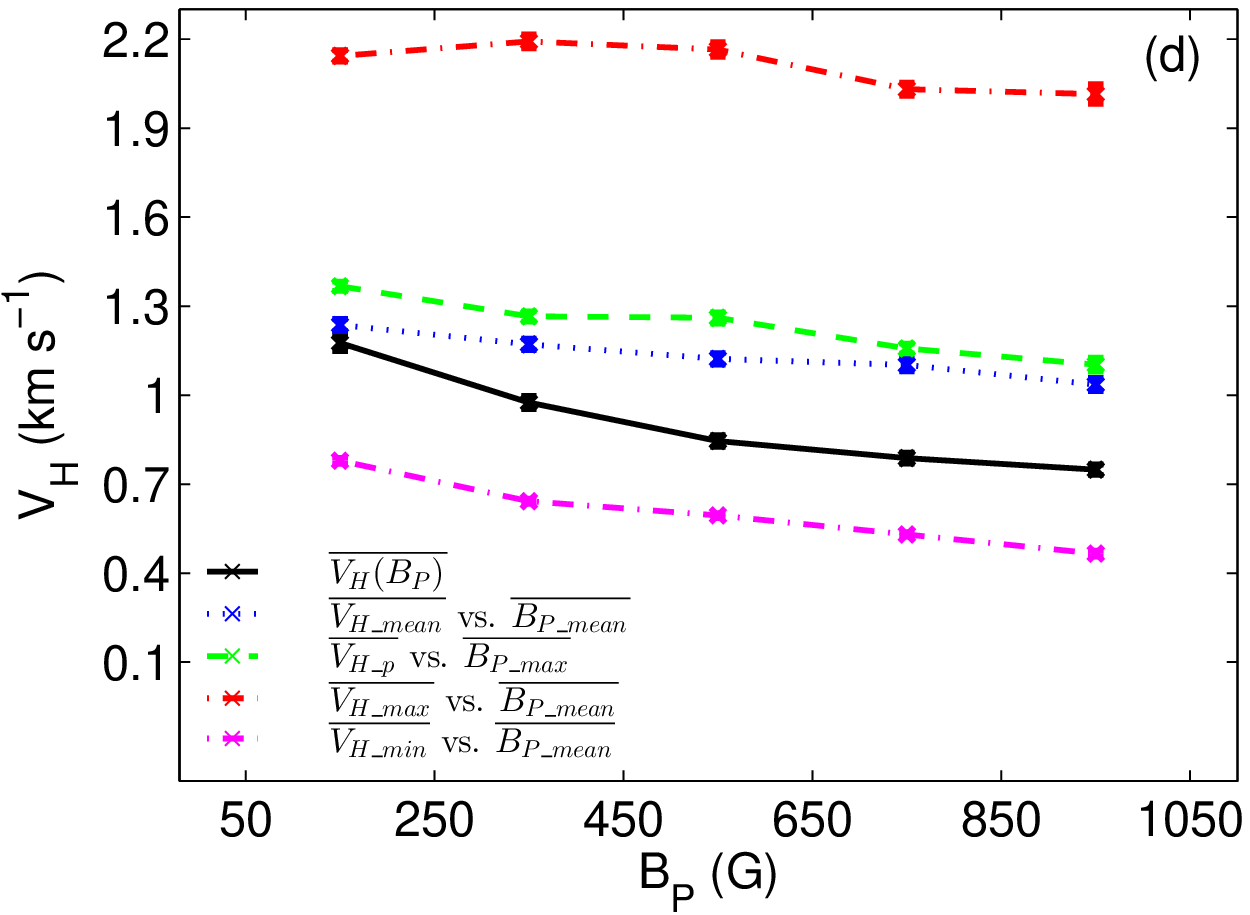}
\caption{(a) The two-dimensional histogram of $B_P$ and $V_H$ of the GBPs in all frames of the time-series. The $x$, $y$ and $z$ coordinates are $B_P$ ($\rm kG$), $V_H$ ($\rm km$ $\rm s^{-1}$) and the normalized frequency ($N/N_{max}$, where $N$ is the number of each bin, $N_{max}$ is the maximum number of $N$). (b) The Rayleigh curve fitted lines of the $V_H$ distributions of the five $B_P$ groups. (c) The $\sigma$ values and the standard errors of the Rayleigh distributions of the five $B_P$ groups (black dashed line), and the trend of the $\sigma$ values is fitted with an exponential function (green solid line). The $V_H$ values with their standard errors of the five $B_P$ groups (blue dashed line) and the exponential fitted line (red solid line). (d) The $B_P$-$V_H$ relations of the GBPs, which the cross points indicate the mean values and the error bars indicate the associated standard errors. The black solid line depicts the $V_H(B_P)$ relation of (a), and the others are obtained via averaging over evolutionary GBP tracks. They are the $V_{H\_mean}$ \textit{vs.} $B_{P\_mean}$ relation (blue dotted line), the $V_{H\_p}$ \textit{vs.} $B_{P\_max}$ relation (green dashed line), the $V_{H\_max}$ \textit{vs.} $B_{P\_mean}$ relation (red dash-dot line) and the $V_{H\_min}$ \textit{vs.} $B_{P\_mean}$ relation (pink dash-dot line), respectively. An overline indicates that the parameter value is averaged for each of the magnetic field strength subgroups.}
    \label{fig:figure3}
\end{figure}

Between every two sequential frames, we averaged the $B_P$ values and calculated the horizontal velocity, $V_H$, by the barycenters of the GBPs. Then, all $V_H$ values of the time-series were categorized into five groups based on the averaged $B_P$ values. Figure~\ref{fig:figure3}(a) shows the distributions of the $V_H$ values of the five $B_P$ groups using a two-dimensional histogram. The $x$, $y$ and $z$ coordinates are $B_P$, $V_H$ and the normalized frequency, respectively.
The $V_H$ distribution of each $B_P$ group is fitted well to a Rayleigh function respectively. Figure~\ref{fig:figure3}(b) shows the curve fitted lines of the Rayleigh functions. A Rayleigh function is described as
\begin{equation}\label{eqa1}
\frac{x}{\sigma^{2}}e^{-\frac{x^{2}}{2\sigma^{2}}},
\end{equation}
where $\sigma$ is the scale parameter of the distribution. Figure~\ref{fig:figure3}(c) shows the $\sigma$ values and the standard errors of the five $B_P$ groups (black dashed line). It can be seen that the $\sigma$ values decrease with the increase of the $B_P$ values. The trend of the $\sigma$ values is fitted well to an exponential function (green solid line). The mean and standard deviation of a Rayleigh distribution are expressed as $\sigma\sqrt{\pi/2}$ and $\sigma\sqrt{(4-\pi)/\sigma^{2}}$ respectively. The standard error describes the possible variation of the mean, or the error of the estimated mean value. It can be calculated by dividing the standard deviation by the square root of the number of measurements. Table~\ref{Tab:tab2} lists the mean $V_H$ values and the standard error values of the five $B_P$ groups. The mean $V_H$ value with standard error is 1.18$\pm$0.03\,$\rm km$ $\rm s^{-1}$ of the first $B_P$ group; then, it decreases with the increase of the $B_P$ values; in the final group reaching 0.75$\pm$0.02\,$\rm km$ $\rm s^{-1}$. Figure~\ref{fig:figure3}(c) also shows the $B_P$ values with their standard errors of the five $B_P$ groups (blue dashed line) and the exponential fitted line (red solid line). The $V_H$($B_P$) relation is approximated by an exponential function given as $ae^{bx}+c$; the fitting yields 0.73$\pm$0.01\,$\rm km$ $\rm s^{-1}$, -2.83$\pm$0.19\,$\rm kG^{-1}$ and 0.70$\pm$0.01\,$\rm km$ $\rm s^{-1}$ for the parameters of $a$, $b$ and $c$, respectively.

\begin{table}
\begin{center}
\caption[]{The GBP Properties of the Five $B_P$ Groups}\label{Tab:tab2}
\begin{threeparttable}
  \begin{tabular}{lccccc}
  \hline\noalign{\smallskip}
  Groups (G)    & [50, 250)\tnote{1} & [250, 450) & [450, 650) & [650, 850) & [850, 1050]\tnote{2}  \\
  \hline\noalign{\smallskip}
Mean $V_H$ ($\rm km$ $\rm s^{-1}$)     &1.18$\pm$0.03	&0.98$\pm$0.02 &0.85$\pm$0.02   &0.79$\pm$0.02  &0.75$\pm$0.02 \\
Number of stationary GBPs               &2353 	&2496 	&2971 	&3797 	&1731\\
Number of non-stationary GBPs           &1441 	&1096 	&863 	&860 	&454\\
Proportion of non-stationary GBPs       & 38\%  & 31\%   & 23\%   &18\%   &21\%\\
Mean $M_R$                             &1.73$\pm$0.02     &1.68$\pm$0.02     &1.49$\pm$0.02     &1.41$\pm$0.01     &1.40$\pm$0.02\\
Maximum $M_R$                              &7.71  &6.42  &4.87  &5.26  &3.47\\
Mean $M_T$                             &0.82$\pm$0.01  &0.73$\pm$0.01  &0.71$\pm$0.01  &0.71$\pm$0.01  &0.70$\pm$0.01\\
  \noalign{\smallskip}\hline
\end{tabular}
\begin{tablenotes}
        \footnotesize
        \item[1] [$a$, $b$): $a\leq x < b$
        \item[2] [$a$, $b$]: $a\leq x \leq b$
 \end{tablenotes}
\end{threeparttable}
\end{center}
\end{table}

We obtained similar results after analyzing the relations using another parameters obtained over evolutionary GBP tracks. The descriptions of the parameters are listed in Table~\ref{Tab:tab3}. Figure~\ref{fig:figure3}(d) shows the relations of $V_{H\_mean}$ \textit{vs.} $B_{P\_mean}$, of $V_{H\_p}$ \textit{vs.} $B_{P\_max}$, of $V_{H\_max}$ \textit{vs.} $B_{P\_mean}$ and of $V_{H\_min}$ \textit{vs.} $B_{P\_mean}$, respectively. When a parameter is depicted with an overline, it has the meaning that this value is averaged for each of the magnetic field strength subgroups. These relations indicate different evolutionary aspects of the evolution of GBPs, such as the mean, maximum and minimum velocity during its lifetime, the velocity at its maximum field strength moment during its lifetime. Aside from different slopes, all the lines show decrease trend with the increase of the magnetic field strengths.

All of the relations between GBP magnetic field strengths and horizontal velocities suggest that the weakly magnetized GBPs move fast, while the strongly magnetized ones move slowly. Namely, magnetic elements showing strong magnetic field strengths move more slowly than weak ones. The result further advances the early conclusion of \citet{Schrijver1996Dynamics}, who found that the horizontal velocity of Ca \textsc{ii} K chromospheric bright points decreases with the increase of their magnetic flux densities. Both relations are consistent with each other, which also implies that there is a close relationship between photospheric bright points and chromospheric bright points.

\begin{table}
\begin{center}
\caption[]{The parameters displayed in Figure~\ref{fig:figure3}(d)}\label{Tab:tab3}
  \begin{tabular}{ll}
  \hline\noalign{\smallskip}
  Parameter & Description   \\
  \hline\noalign{\smallskip}
$V_{H\_mean}$           &the mean $V_H$ value during a GBP lifetime\\
$V_{H\_max}$            &the maximum $V_H$ value during a GBP lifetime\\
$V_{H\_min}$            &the minimum $V_H$ value during a GBP lifetime\\
$V_{H\_p}$              &the $V_H$ value at the maximum $B_P$ moment during a GBP lifetime\\
$B_{P\_mean}$           &the mean $B_P$ value during a GBP lifetime\\
$B_{P\_max}$            &the maximum $B_P$ value during a GBP lifetime\\
\noalign{\smallskip}\hline
\end{tabular}
\end{center}
\end{table}

\subsection{The relation between $B_P$ and motion range}
The motion ranges of all GBPs were measured using the definition by \citet{Bodnarova2014On} and \citet{Yang2015Characterizing}. An index of motion range, $M_R$, is defined as $M_R=\sqrt{(X_{max}-X_{min})^{2}+(Y_{max}-Y_{min})^{2}}/r_{max}$, where $X_{max}$ and $X_{min}$ are the maximum and minimum coordinates of the path of a GBP in the $x$ axis, and $Y_{max}$ and $Y_{min}$ are in the $y$ axis; $r_{max}$ is the radius of the circle which corresponds to the maximum size of the GBP during its lifetime. Then, we categorized all GBPs into stationary and non-stationary according to their $M_R$ values. A GBP is regarded as a stationary one if its motion range is smaller than its own maximum radius during its lifetime ($M_R <$ 1). Otherwise, it is a non-stationary one ($M_R \geq$ 1). Stationary GBPs account for 74\% of all GBPs in this data set. It implies that the majority of GBPs are jostled in a limited area, while the minority are pushed by granules far away. The proportion of stationary GBPs is higher than that reported by \citet{Bodnarova2014On} and \citet{Yang2015Characterizing}. They both indicated that there are about 50\% stationary GBPs in a QS region. We speculated that the difference comes from the magnetized environments that the GBPs are embedded in.

\begin{figure}[ht]
 \centering
\includegraphics[width=0.45\textwidth]{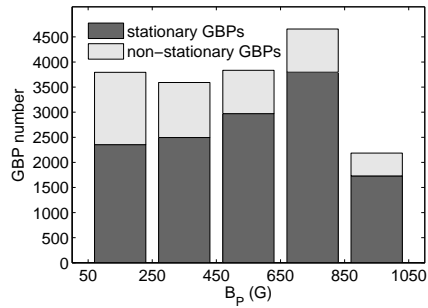}
\caption{The distribution of the stationary and non-stationary GBPs. The proportions of stationary GBPs increase from 62\% to 79\% with the increase of the $B_P$ values.}
    \label{fig:figure4}
\end{figure}

Figure~\ref{fig:figure4} shows the distribution of the stationary and non-stationary GBPs in each $B_P$ group. Their numbers are also listed in Table~\ref{Tab:tab2}. The proportion of stationary GBPs increases from 62\% to 79\% with the increase of the {$B_P$} values. This implies that the majority of GBPs are localised, especially the very strongly magnetized GBPs.

\begin{figure}[ht]
 \centering
\includegraphics[width=0.45\textwidth]{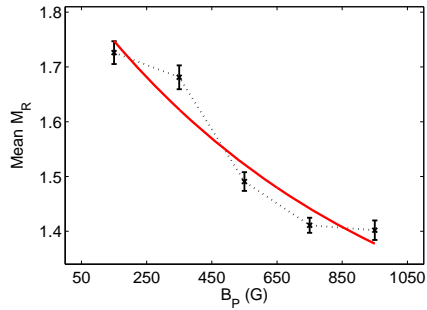}
\caption{The $M_R$($B_P$) relation, which the crosses indicate the mean $M_R$ values and the error bars indicate the associated standard errors of the non-stationary GBPs in each $B_P$ group. The relation is fitted with an exponential function (red solid line). }
    \label{fig:figure5}
\end{figure}

Figure~\ref{fig:figure5} shows the $M_R(B_P)$ relation, which the crosses indicate the mean $M_R$ values and the error bars indicate the associated standard errors of the non-stationary GBPs in each $B_P$ group. The mean $M_R$ value with standard error is 1.73$\pm$0.02 in the first $B_P$ group; it decreases with the increase of the $B_P$ values, reaching 1.40$\pm$0.02. The relation is approximated by an exponential function given as $ae^{bx}+c$; the fitting yields 0.74$\pm$0.54, -1.11$\pm$1.69\,$\rm kG^{-1}$ and 1.12$\pm$0.64 for the parameters of $a$, $b$ and $c$, respectively.

Furthermore, the maximum $M_R$ values of the five $B_P$ groups were determined (see Table~\ref{Tab:tab2}). Similarly to the mean $M_R$ value, the maximum $M_R$ values decrease from 7.71 to 3.47. Our result is consistent with the previous studies that indicated the maximum $M_R$ value of the GBPs in the QS region is about 7 \citep{Bodnarova2014On,Yang2015Characterizing}.
The above results suggest that weakly magnetized GBPs move far away, while strongly magnetized GBPs are more localised.

\subsection{The relation between $B_P$ and motion type}
Putting aside the stationary GBPs, we analyzed the motion type of 4814 non-stationary ones. An index of motion type, $M_T$, is defined as $M_T=D/L$, where $D$ is the displacement of a GBP, defined as $D=\sqrt{(X_{n}-X_{1})^{2}+(Y_{n}-Y_{1})^{2}}$, here $X_{1}$ and $Y_{1}$ are the barycenter position of the start frame during a GBP's lifetime, and $X_{n}$ and $Y_{n}$ are the barycenter position of the final frame; $L$ is the whole path length of the GBP, defined as $L=\sum_{t=1}^{n-1}\sqrt{\triangle X_{t}^{2}+\triangle Y_{t}^{2}}$, here $\triangle X_{t}=X_{t+1}-X_{t}$, $\triangle Y_{t}=Y_{t+1}-Y_{t}$ \citep{Yang2015Characterizing}. The $M_T$ value is a ratio of the displacement of a GBP to its whole path length. According to the definition, the value must be between 0 and 1. If a GBP moves in a nearly straight line, the $M_T$ value will be close to 1. But if it moves in a nearly closed curve, then the $M_T$ value will be close to 0. Please refer the details to \citet{Yang2015Characterizing}. We found the $M_T$ values for half of the detected GBPs are larger than 0.83, and that of 19\% GBPs are less than 0.50. According to the definition of $M_T$, the result implies that most GBPs move in straight or nearly straight lines, and only a few move in intricate or rotary trajectories.

\begin{figure}[ht]
 \centering
\includegraphics[width=0.45\textwidth]{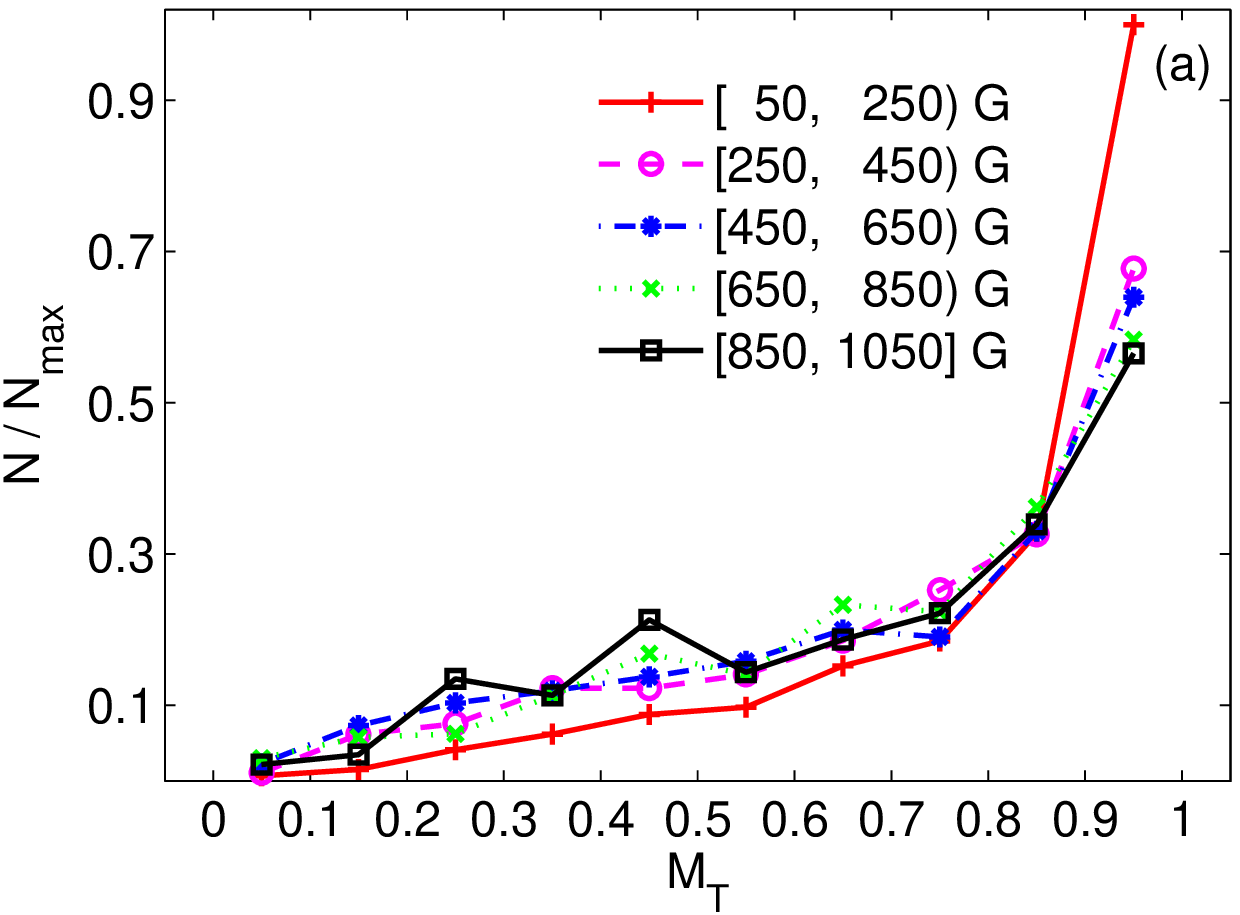}
\includegraphics[width=0.45\textwidth]{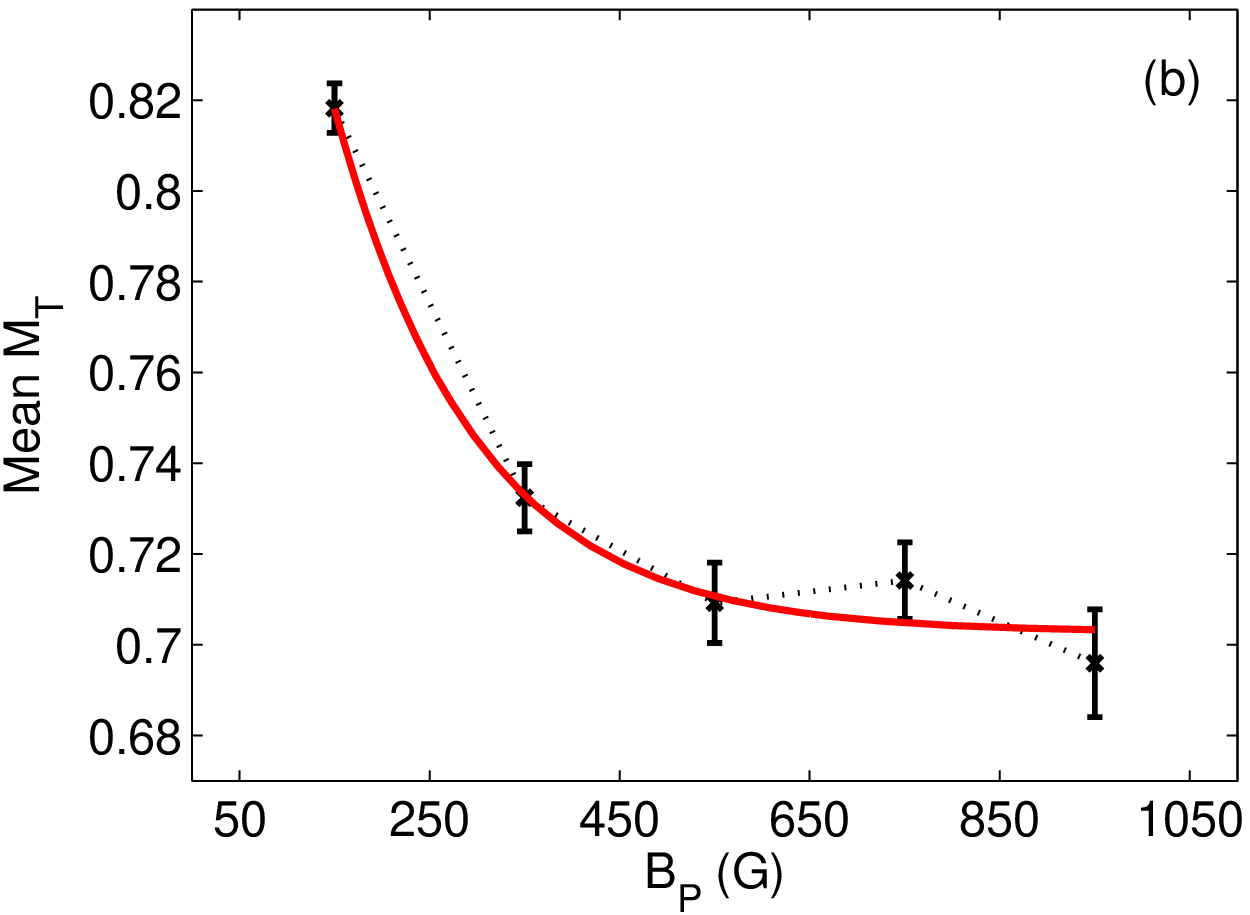}
\caption{(a) the $M_T$ distributions of the non-stationary GBPs of the five $B_P$ groups. (b) the $M_T$($B_P$) relation, which the crosses indicate the mean $M_T$ values and the error bars indicate the associated standard errors of the five $B_P$ groups. The relation is fitted well with an exponential function (red solid line).}
    \label{fig:figure6}
\end{figure}

Figure~\ref{fig:figure6}(a) shows the $M_T$ distributions of the five $B_P$ groups. All of them show an exponential increase, only differ from the rate of increase. The $M_T$ distribution of the strongest $B_P$ group is more flat than that of the weakest $B_P$ group. The weakly magnetized GBPs are found to move in straight lines, while the strongly magnetized GBPs appear to move in intricate or rotary trajectories.

Figure~\ref{fig:figure6}(b) shows the $M_T$($B_P$) relation, which the cross points indicate the mean $M_T$ values and the error bars indicate the associated standard errors of the five $B_P$ groups. The mean $M_T$ value with standard error is 0.82$\pm$0.01 in the first $B_P$ group; then, it decreases with the increase of the $B_P$ values; for the final group reaching 0.70$\pm$0.01. The relation is also approximated by an exponential function given as $ae^{bx}+c$, which yields 0.31$\pm$0.08, -6.69$\pm$1.77\,$\rm kG^{-1}$ and 0.70$\pm$0.01 for the parameters of $a$, $b$ and $c$, respectively.
The trend indicates that very weakly magnetized GBPs generally move in nearly straight lines, and their motions become more and more complicated with the increase of the $B_P$ values.

\section{Discussion}
\label{sec:DISCUSSION}
\subsection{The magnetic field strength distribution of GBPs}

We calibrated the NFI $V/I$ signals using SP longitudinal magnetograms multiplying the filling factors. The longitudinal magnetic field strength of each GBP was extracted from the corresponding calibrated NFI magnetograms by a point-to-point method. The distribution of GBP longitudinal magnetic field strengths is fitted well with a double log-normal function with two peaks at about 140 and 650\,$\rm G$.

The double log-normal distribution is qualitatively in agreement with \citet{Utz2013Magnetic}, who extracted the GBP magnetic field strength at the coordinate of GBP barycenter from the \textit{Hinode} /SOT SP data. They suggested that the two components correspond to the magnetic field strength background signal and the flux tubes with higher magnetic field strengths, respectively. The background distribution was explained by considering of uncollapsed flux tubes, probably in the course of the dissolution, and a low fraction of mis-identified features (non-GBP features). The other component was explained by the flux tubes following the convective collapse, for which \citet{Spruit1979Convection} in a model of convective collapse predicted a magnetic field strength of about 1300\,$\rm G$.

We agree with their interpretation of the obtained two components of the distribution, although the two peak values obtained within the current work are different from theirs. The difference mainly comes from the filling factor, which is the portion of each pixel occupied by the magnetized component. \citet{Utz2013Magnetic} obtained two peaks at about 270 and 1340\,$\rm G$ of the distribution by assuming the filling factor of unity, but we calibrated the NFI $V/I$ signals with the SP longitudinal magnetogram multiplying the filling factors. If we calibrated the NFI $V/I$ signals by assuming the filling factor as unity, the longitudinal magnetic field strength distribution of GBPs would show two peaks at 300 and 1375\,$\rm G$, respectively. However, the measured $V/I$ is proportional to the field strength integrated over the pixel area when the spatial resolution is lower than the size of the magnetic elements. In theory, the lower the spatial resolution, the lower the averaged field strength over the pixel area. This is the main reason why the field strengths of BPs in this study only reach $\rm hG$ values. Our results are consistent with the previous studies. \citet{Berger2001On} found that the average peak flux density of GBPs is about 160\,$\rm G$ using the Lockheed \textit{Solar Optical Universal Polarimeter} (SOUP) filter magnetograms, which the spatial resolution is 0.3$^{\prime\prime}$. \citet{Viticchie2010Imaging} analyzed the spectropolarimetric images acquired at the \textit{Dunn Solar Telescope} (DST) with the \textit{Interferometric Bidimensional Spectrometer} (IBIS), which the pixel scale is 0.18$^{\prime\prime}$. They got the peak value for distribution of the average flux density in BPs is approximatively 100--150\,$\rm G$ with a tail up to 800\,$\rm G$ using the center-of-gravity (COG) method. \citet{Riethmuller2014Comparison} compared high-resolution \textit{Sunrise} data in three spectral bands with three-dimensional radiative MHD simulations. They found that the BPs in the observation data only show 540\,$\rm G$ field strengths based on the weak field approximation, while all magnetic BPs in the simulations are associated with $\rm kG$ fields. On the other hand, the inverted magnetograms from the spectropolarimetric data with the filling factors as a free parameter can show $\rm kG$ field strengths of BPs \citep{Beck2007Magnetic,Viticchie2010Imaging,Utz2013Magnetic}.

The theoretical model indicates only \,$\rm kG$ magnetic field concentrations appear bright in the photosphere \citep{Spruit1979Convection}. \citet{Riethmuller2014Comparison} indicate that all magnetic BPs identified on the basis of their brightness properties in the simulations are associated with $\rm kG$ fields, with the average field strength at constant optical depth log($\tau$)$=$ 0 being 1760\,$\rm G$, which drops to 1070\,$\rm G$ at log($\tau$)$= -$2. \citet{Criscuoli2014statistical} applied 3D magneto-hydrodynamic simulations and then proposed that the field strength distribution of G-band bright features shows two peaks. One at about 1.5\,$\rm kG$ corresponds to magnetic features and one at 1\,$\rm hG$ mostly corresponds to bright granules. However, the majority of GBPs cannot reach \,$\rm kG$ field strengths in this study (even taking the filling factor as unity), as well as the previous works cited above. The author \citet{vanBallegooijen1985Electric} suggested early that the plausible explanation is randomly occurring inclinations of small flux elements render the longitudinal field below the detection limit at random times. \citet{SanchezAlmeida2000Physical} elucidated the micro-structured magnetic atmospheres hypothesis that the highest field strength filaments have the lowest opacity and thus the lowest measured polarization signal, while the observed flux concentration is in essence a density-weighted average that largely favors the weak-field and nonmagnetic filaments. \citet{Berger2001On} discussed that this is mainly due to misidentification of granulation brightening and insufficient magnetogram flux sensitivity. \citet{Criscuoli2014statistical} explained that it is probably caused by insufficient spatial resolution of magnetograms. \citet{Riethmuller2014Comparison} indicated that the weakly polarized signals of the most BPs in the observational data can be partly explained by a combination of thermal weakening of the temperature sensitive Fe \textsc{i} 525.0 $\rm nm$ line, spatial smearing due to residual pointing jitter, and instrumental stray light.

To verify whether the weaker GBPs whose magnetic field strengths are less than about 400\,$\rm G$ (the valley of the distribution) are bright points in intergranular lanes or granulation brightening, we have checked them in the G-band images by visual identification. The result shows that the majority of them are indeed bright points in intergranular lanes, and a few are difficult to distinguish them by eyes. This confirms that the observed flux concentration is a density-weighted average that largely favors the weak-field with a small filling factor. Therefore, we agree that the GBPs showing low magnetic field strengths are probably due to instrumental effects, e.g., insufficient magnetogram flux sensitivity, or insufficient spatial resolution, or PSF, or measurement noise. The measurement noise is mainly composed of the photon noise and the noise induced by the instruments. In general, the method of re-binning or exposure time increase can improve signal to noise ratio (SNR), however, they will result in reducing the spatial or temporal resolution. Higher spatially as well as temporally resolved magnetograms are still needed to reveal the fine-structure of small-scale magnetic elements and the detailed relationship between GBP dynamics and the corresponding magnetic element in the future.

\subsection{The dynamics \textit{vs.} $B_P$ relation}

The relations of $V_H(B_P)$, $M_R(B_P)$ and $M_T(B_P)$ describe the dynamics of GBPs with different magnetic field strengths, in terms of the horizontal velocity, motion range and motion type. We found all of these decrease with the increase of the GBP magnetic field strengths, which can be described by exponential functions. It implies that the magnetic elements showing weak magnetic field move fast and far in nearly straight lines, while the magnetic elements showing strong magnetic field move slowly within a small area along complicated paths.

To make sure whether the peak magnetic field strength of GBPs is a good proxy for the change in the dynamics, we cross-checked the correlations of GBP dynamics and some other GBP properties for the very strongly magnetized GBPs, in terms of the total magnetic fluxes, the size and the lifetime of GBPs.

\begin{figure}[ht]
 \centering
\includegraphics[width=0.45\textwidth]{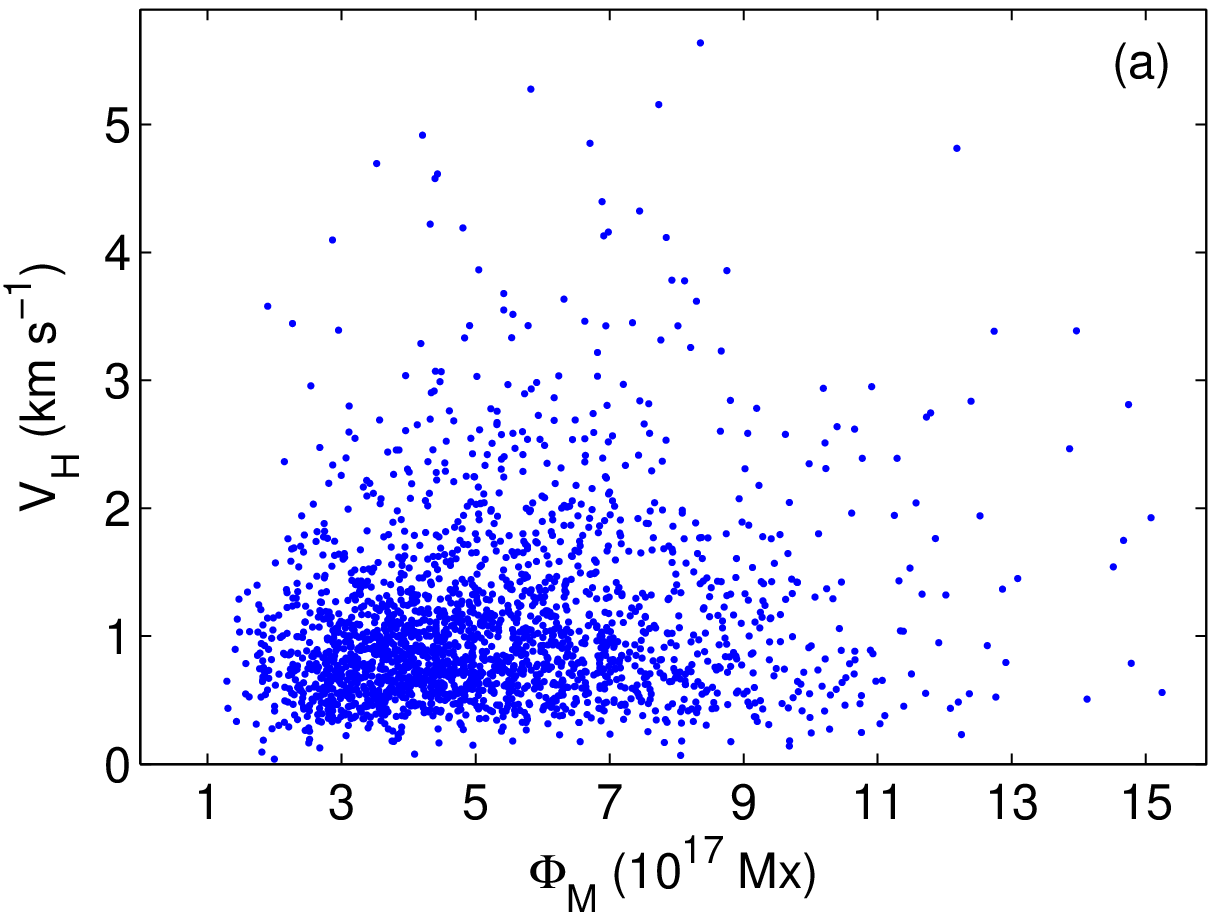}
\includegraphics[width=0.447\textwidth]{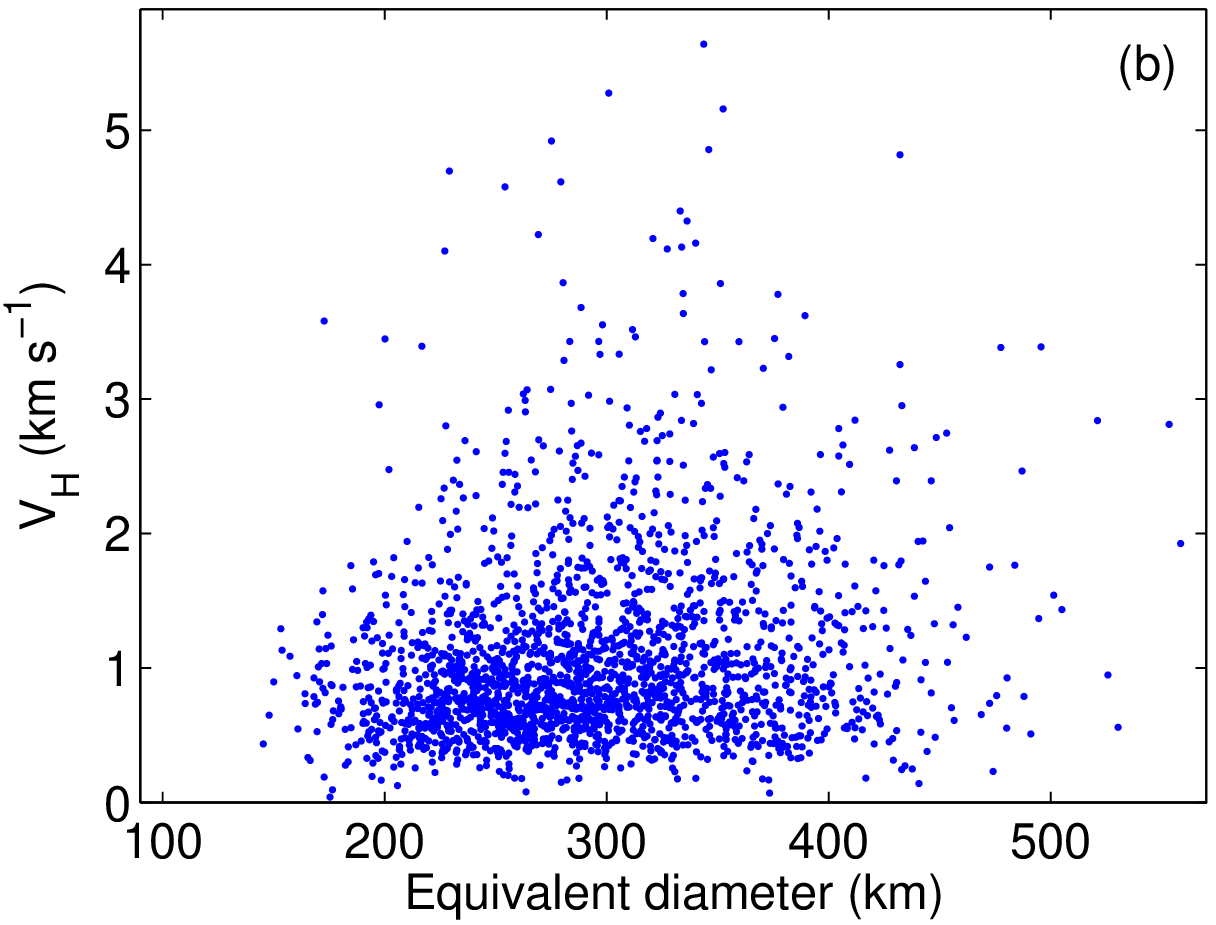}
\caption{(a) Scatter plot of $\Phi_M$ \textit{vs.} $V_H$. (b) Scatter plot of the equivalent diameter \textit{vs.} $V_H$. Both of them are plotted for the strongest $B_P$ group.}
    \label{fig:figure7}
\end{figure}

The first one is the correlation between $V_H$ and the total magnetic fluxes of GBP for the $B_P$ group from 850 to 1050\,$\rm G$ (very strongly magnetized GBPs). The total magnetic fluxes of a GBP, $\Phi_M$, is the integral of the magnetic field strengths over its area. Figure~\ref{fig:figure7}(a) shows a scatter plot of $\Phi_M$ \textit{vs.} $V_H$. It can be seen that there is no obvious relation between them. We also calculated the correlation degree between them, and obtained a value of 0.16. This implies that the horizontal velocity is unlikely to depend on the total magnetic flux of GBPs.
Then, the correlation of $V_H$ and the size of GBPs was checked. Figure~\ref{fig:figure7}(b) shows a scatter plot of the equivalent diameter \textit{vs.} $V_H$ for the strongest $B_P$ group. The same as in Figure~\ref{fig:figure7}(a), there is no obvious relation between them, having a correlation degree of 0.17.

We also repeated these checking for the other $B_P$ groups, and obtained similar results. These findings suggest that the peak magnetic field strength ($B_P$) of a GBP is indeed the most important parameter for restraining and inhibiting the movement of the features while the size and/or the total flux of the feature does not play an essential role.

\begin{figure}[ht]
 \centering
\includegraphics[width=0.32\textwidth]{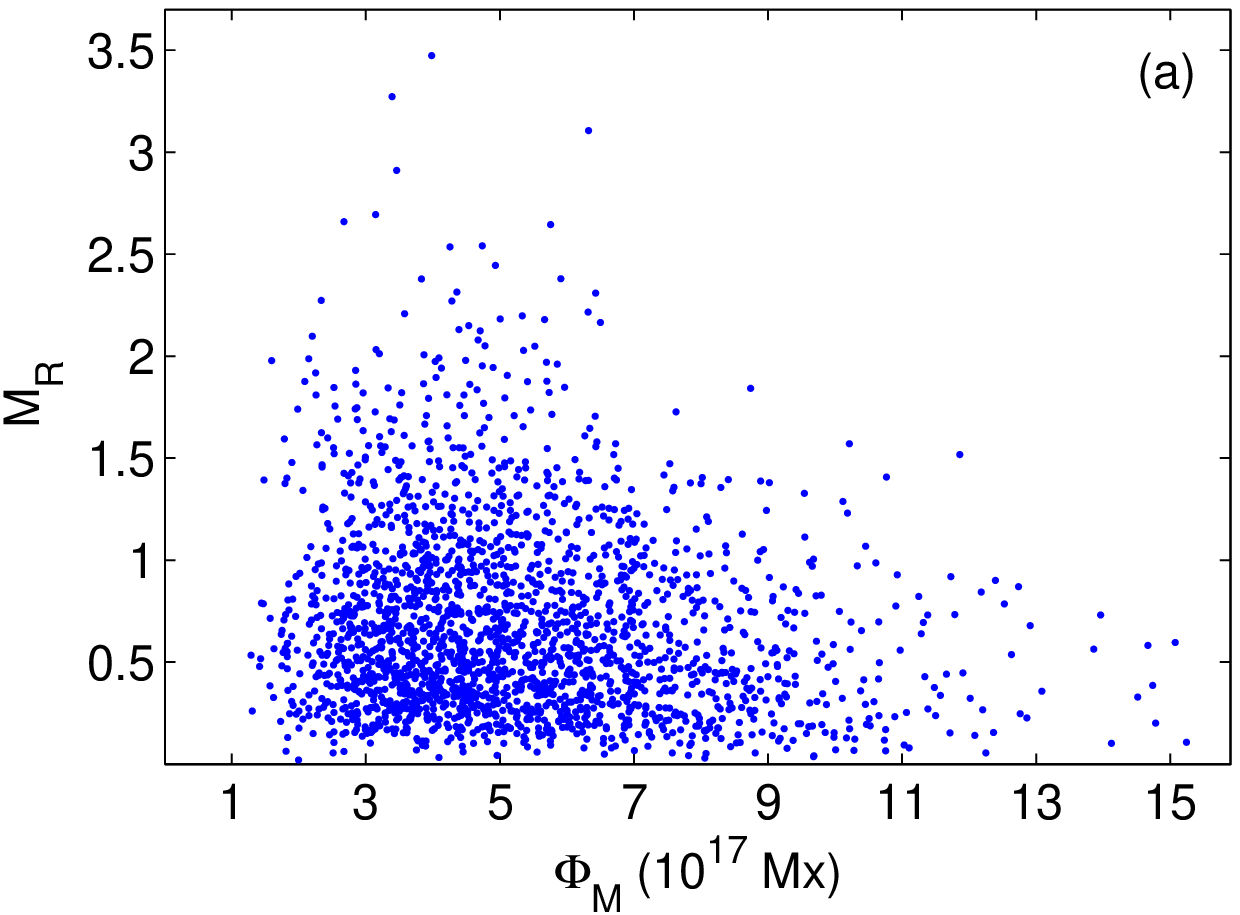}
\includegraphics[width=0.32\textwidth]{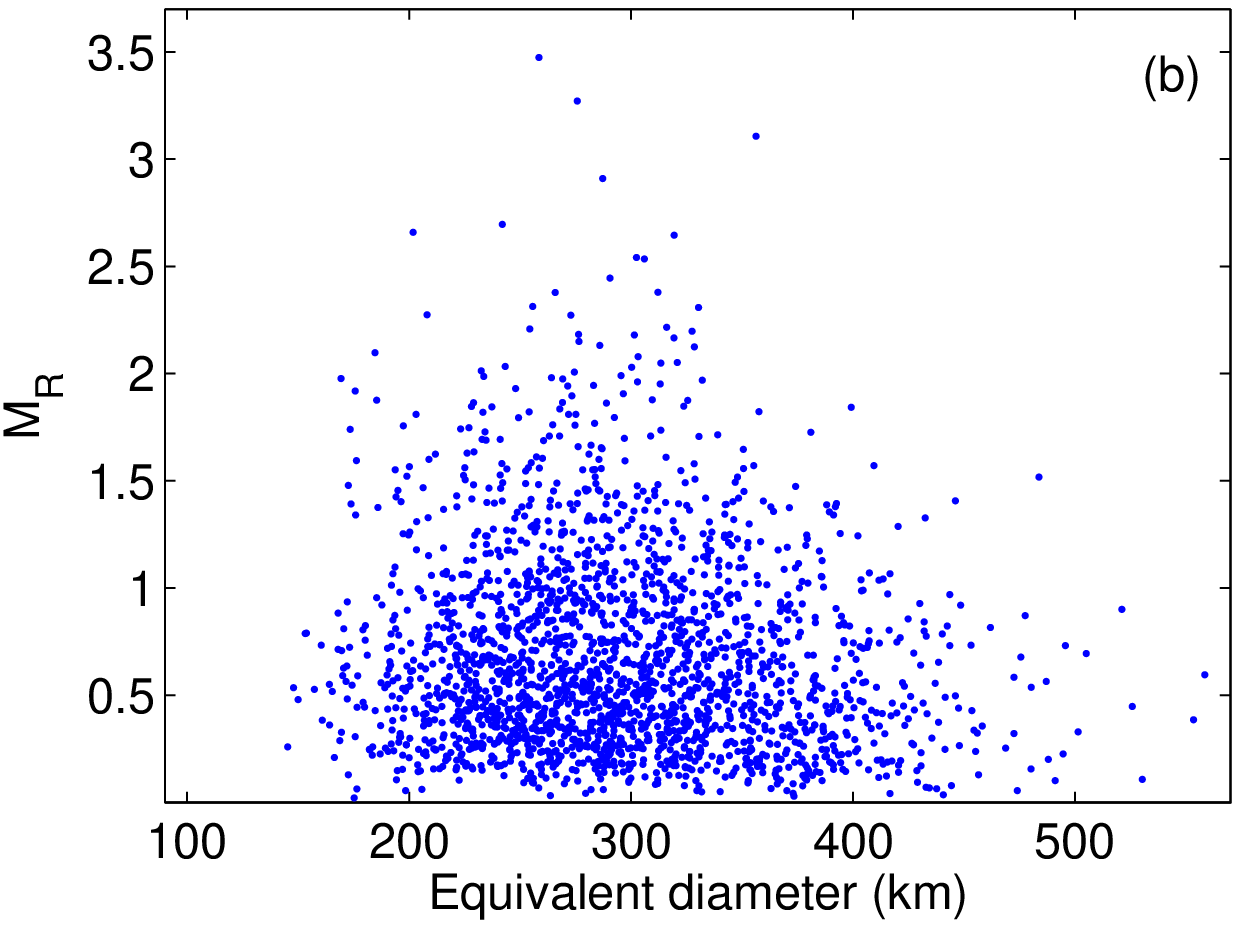}
\includegraphics[width=0.32\textwidth]{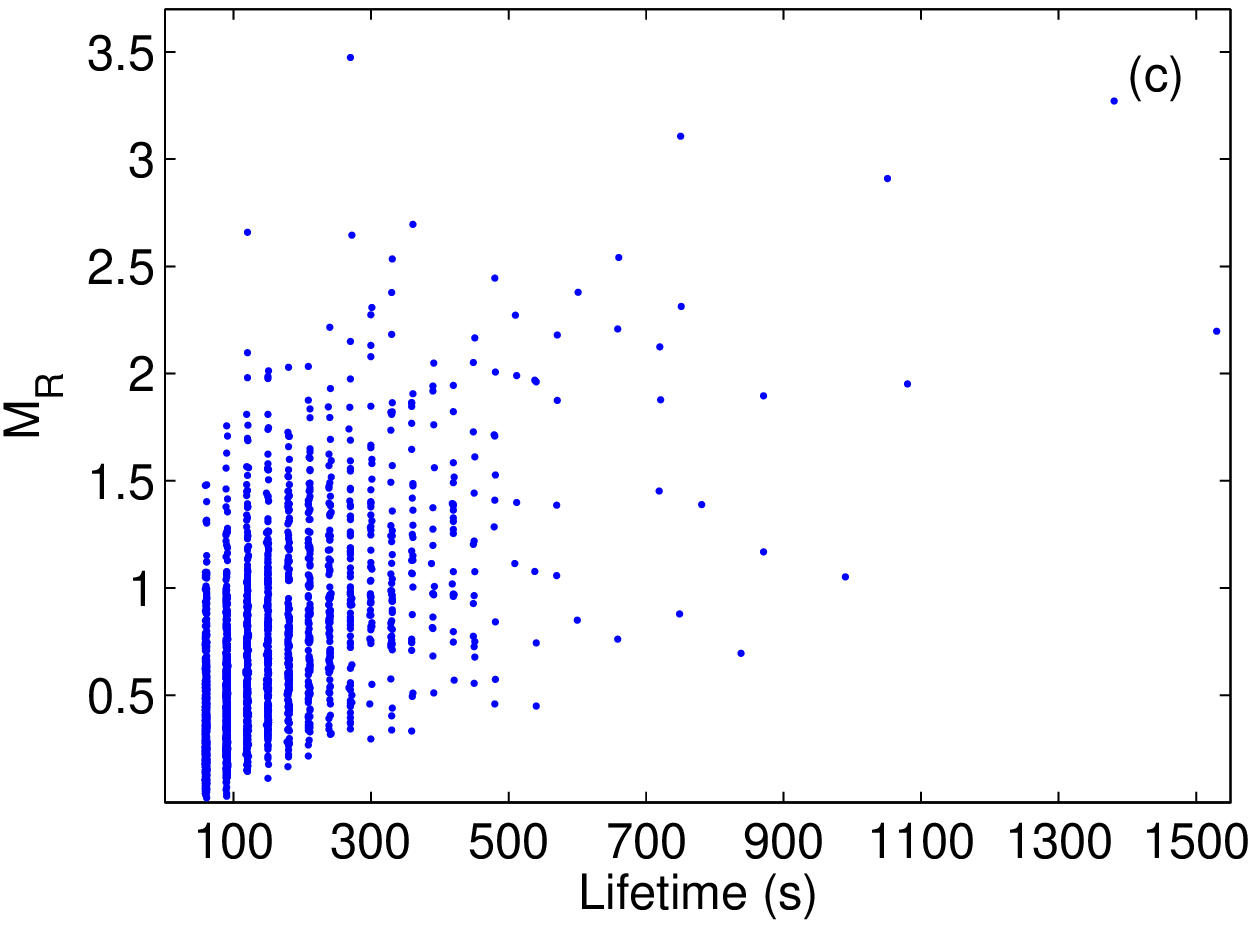}
\caption{(a) Scatter plot of $\Phi_M$ \textit{vs.} $M_R$. (b) Scatter plot of the equivalent diameter \textit{vs.} $M_R$. (c) Scatter plot of the lifetime \textit{vs.} $M_R$. All of them are plotted for the strongest $B_P$ group.}
    \label{fig:figure8}
\end{figure}

In terms of the motion range, besides the total magnetic fluxes and the size of GBPs, the lifetime has been checked. Figure~\ref{fig:figure8} shows scatter plots of $\Phi_M$ \textit{vs.} $M_R$, the equivalent diameter \textit{vs.} $M_R$ and the lifetime \textit{vs.} $M_R$ for the strongest $B_P$ group. The correlation degrees between them are -0.09, -0.15 and 0.58, respectively. There is no correlation with the total magnetic flux and the size, however, there is a certain correlation with the lifetime. Indeed, a short-lived GBP has not the time to move far enough to cross over its own radius in distance. Conversely, it may not always be the case that a long-lived GBP moves far away. The distance a GBP can move does not only depend on the lifetime, but also on the motion type. A long-lived GBP can move far away only if it moves in a straight line, otherwise it will move a short range in an intricate or rotary trajectory. Therefore, the long lifetime is a necessary but not a sufficient condition for the large motion range of a GBP.

\begin{figure}[ht]
 \centering
\includegraphics[width=0.32\textwidth]{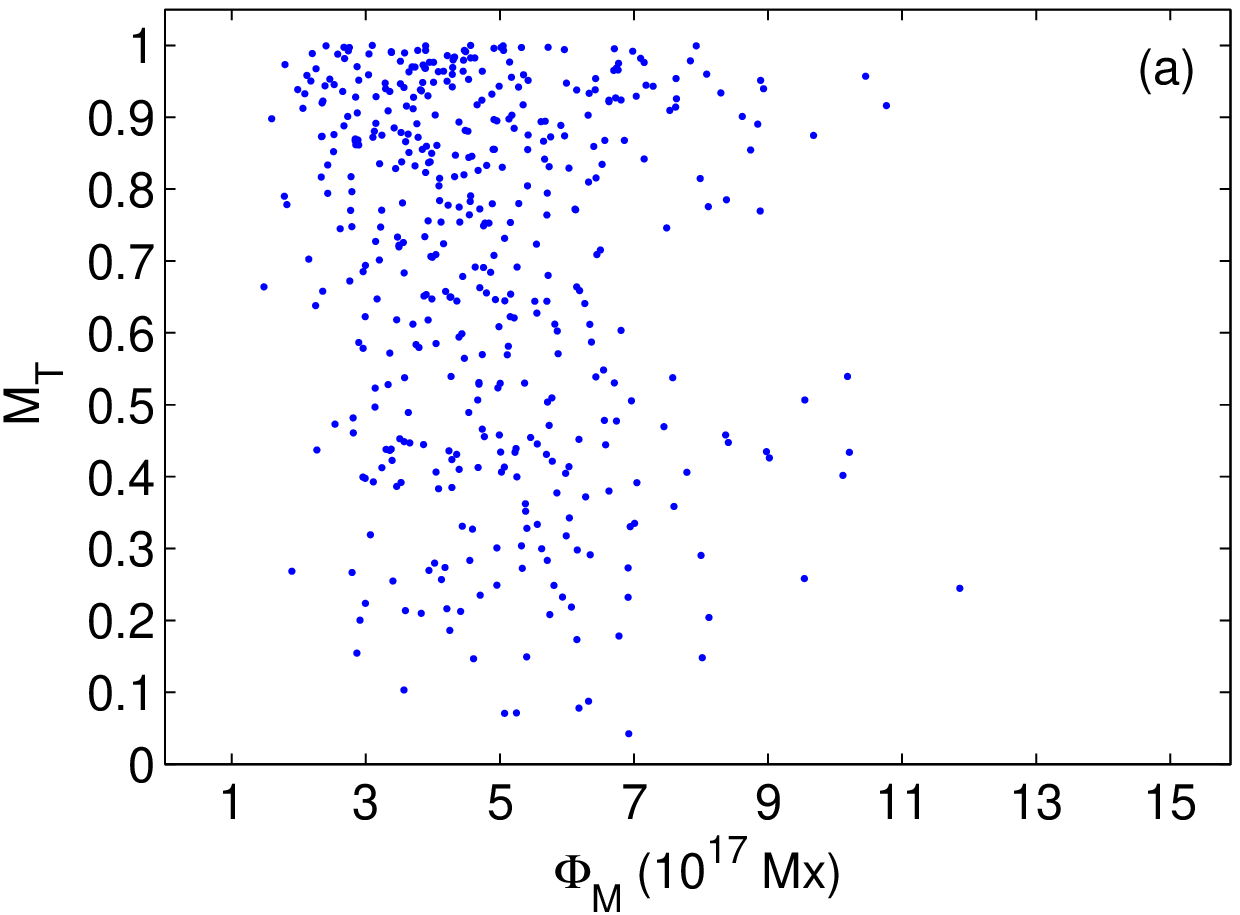}
\includegraphics[width=0.32\textwidth]{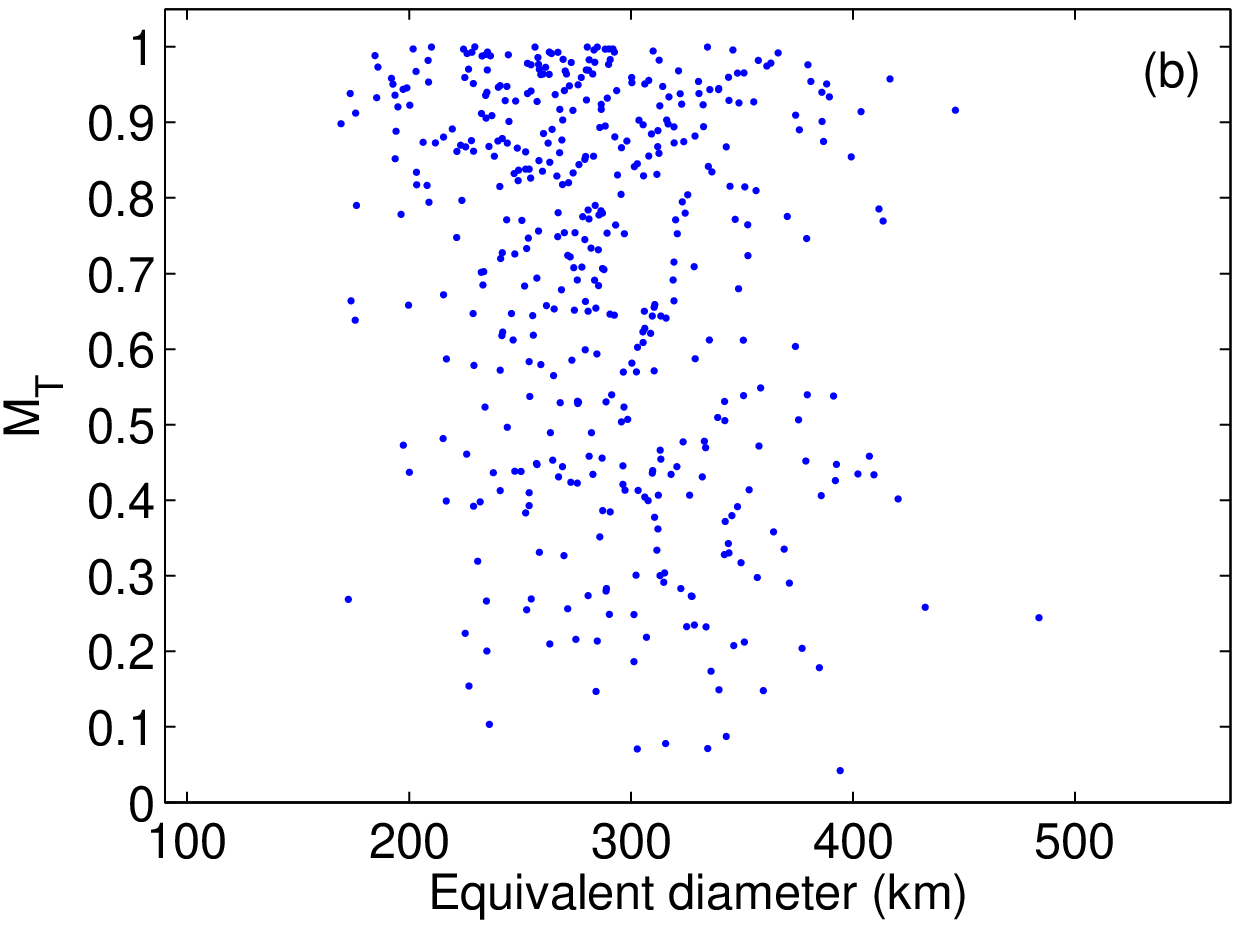}
\includegraphics[width=0.32\textwidth]{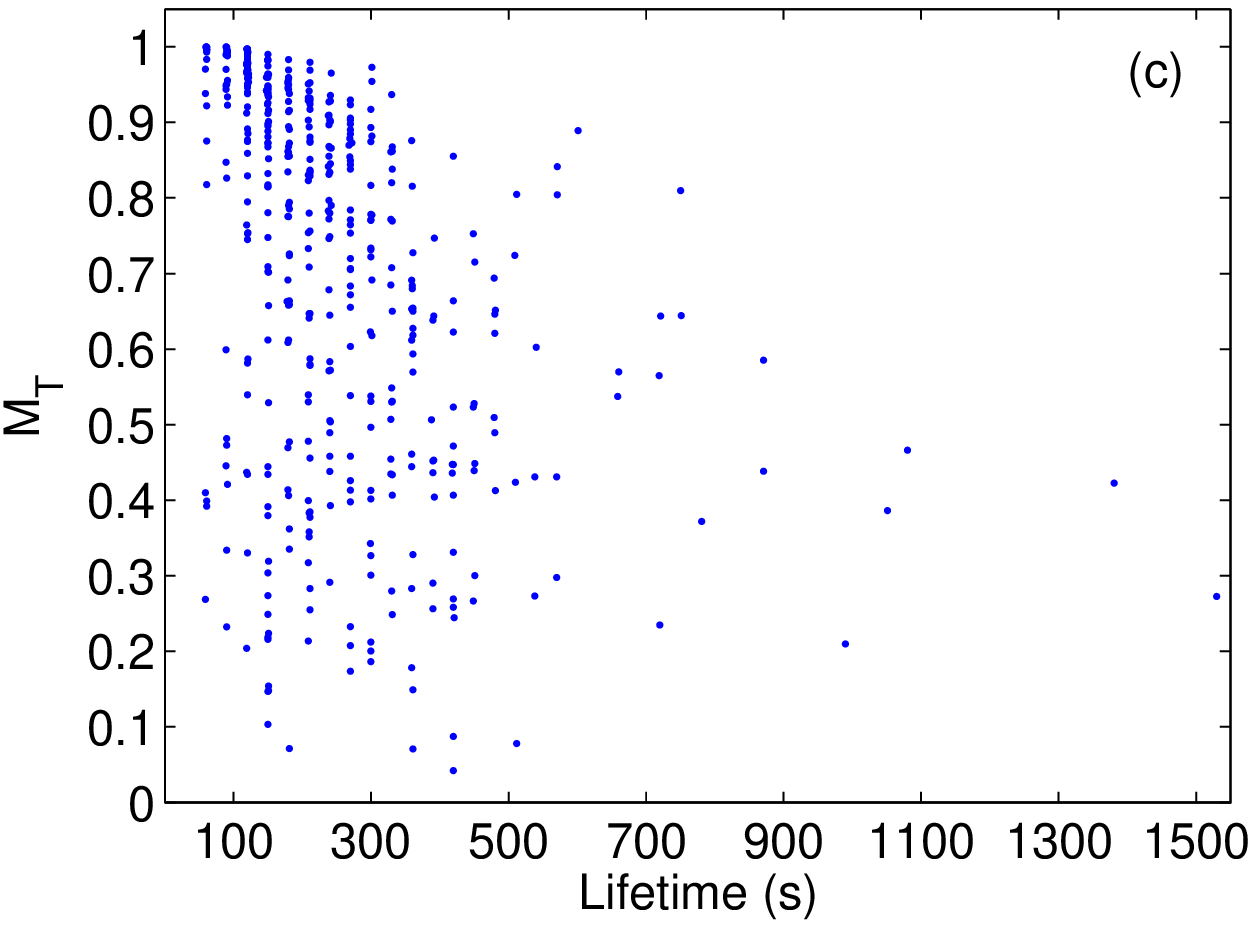}
\caption{(a) Scatter plot of $\Phi_M$ \textit{vs.} $M_T$. (b) Scatter plot of the equivalent diameter \textit{vs.} $M_T$. (c) Scatter plot of the lifetime \textit{vs.} $M_T$. All of them are plotted for the strongest $B_P$ group.}
    \label{fig:figure9}
\end{figure}

Figure~\ref{fig:figure9}(a) and (b) shows scatter plots of $\Phi_M$ \textit{vs.} $M_T$, the equivalent diameter \textit{vs.} $M_T$ for the strongest $B_P$ group, respectively. The correlation degrees between them are -0.15, -0.22, respectively. Both of them show a weak or no correlation with the total magnetic fluxes and the size. Figure~\ref{fig:figure9}(c) shows a scatter plot the lifetime \textit{vs.} $M_T$, with a correlation degree of -0.38. It can be seen that the long-lived GBPs do not move in straight lines in Figure~\ref{fig:figure9}(c), though the correlation degree is not high. The mean $M_T$ value is 0.51 for the very strongly magnetized GBPs with lifetimes are larger than 500\,$\rm s$. This result further explains the above discussion on the motion range. The motion type is the main reason why a strongly magnetized GBP moves a short range even it lives long.

We are unable to affirm which property of GBPs is the determinant factor of their dynamics. However, our analysis suggests that the peak magnetic field strength of GBPs is a good proxy for that. GBPs with strong magnetic field seem to have inhibited movement, in terms of the horizontal velocity, motion range and motion type.

\section{Conclusions}
\label{sec:CONCLUSION}
GBPs show a strong spatial correlation with magnetic flux concentrations, which allows the dynamics of magnetic features to be studied in higher resolution observations. The high spatial and temporal resolution G-band data and NFI Stokes $I$ and $V$ data with \textit{Hinode} /SOT were used to explore the relationship between the dynamics of GBPs and their corresponding longitudinal magnetic field strengths. The NFI images are co-spatial and nearly co-temporal with the G-band images, and have a large FOV. It provides a good opportunity for extracting the simultaneous longitudinal magnetic field strength of each GBP during its lifetime. We concentrated on the dynamics of small-scale magnetic elements with different longitudinal magnetic field strengths, which differs from most previous works that analyzed the dynamics of small-scale magnetic elements in different magnetized environments.

We detected 103 023 isolated GBPs in 132 G-band images and tracked 18 494 evolving GBPs. The simultaneous longitudinal magnetic field strength of each GBP during its lifetime has been extracted in the corresponding calibrated NFI magnetograms by a point-to-point method. And then, the GBPs were categorized into five groups according to their longitudinal magnetic field strengths, and their horizontal velocity, their motion range and their motion type of the five groups were also analyzed.

It is found that the GBPs with weak longitudinal magnetic field move fast and far in nearly straight lines; with the increase of the longitudinal magnetic field that the GBPs correspond, they move more slowly in more complicate paths, and seem to stay localized. The different dynamic behaviors of GBPs can be explained by the interaction between magnetic fields and convection. Magnetic elements showing low field strengths are made move fast by convection in nearly straight lines to concentrate; once their field strengths become higher, they usually become concentrated at stagnation points in convective downflow sinks; as a result, they are squeezed in a local place, and move more slowly in erratic paths. It suggests that magnetic elements with strong magnetic field are hardly perturbed by convective motions and \textit{vice versa}. Concentrated magnetic field suppresses efficiently the transport of energy by convection. Their magnetic energy is not negligible compared with the kinetic energy of the gas and therefore the flows cannot perturb them so easily.

This work shows how fast, how far and how intricately the small-scale magnetic elements with different magnetic field strengths move on the solar surface. These results may be useful for the modeling of wave excitation processes and realistic simulations. In a future study we will focus on the longitudinal movements associated with the magnetic field strengths of GBPs.

%
 \begin{acks}
The authors are grateful to the anonymous referee for constructive comments and detailed suggestions to this manuscript. The authors are grateful to the support received from the National Natural Science Foundation of China (No: 11303011, 11573012, 11463003, U1231205), Open Research Program of the Key Laboratory of Solar Activity of the Chinese Academy of Sciences (No: KLSA201414, KLSA201505). The authors are grateful to the \textit{Hinode} team for the possibility to use their data. \textit{Hinode} is a Japanese mission developed and launched by ISAS/JAXA, collaborating with NAOJ as a domestic partner, NASA and STFC (UK) as international partners. Scientific operation of the \textit{Hinode} mission is conducted by the \textit{Hinode} science team organized at ISAS/JAXA. This team mainly consists of scientists from institutes in the partner countries. Support for the post launch operation is provided by JAXA and NAOJ (Japan), STFC (U.K.), NASA (U.S.A.), ESA, and NSC (Norway).
\end{acks}

%
%

 \bibliographystyle{spr-mp-sola}
 \bibliography{gbp}

\end{article}
\end{document}